\documentclass[procedia]{easychair}

\usepackage{doc}
\usepackage{makeidx}
\usepackage{amsmath,amssymb}
\usepackage{enumerate}
\usepackage{bm}
\usepackage{color}
\usepackage{braket} 
\usepackage[at]{easylist}
\usepackage{graphicx}

\def\procediaConference{20th International Conference of Magnetism (ICM 2015)}

\newcommand{\calH}{\mathcal{H}}
\newcommand{\calO}{\mathcal{O}}

\newcommand{\eff}{\mathrm{eff}}

\title{Laser-induced phase transitions of \\ topological Kondo insulators}

\titlerunning{Laser-induced phase transitions of topological Kondo insulators}

\author{
    Kazuaki Takasan\thanks{E-mail address : \texttt{takasan@scphys.kyoto-u.ac.jp}} \and Masaya Nakagawa \and Norio Kawakami
}

\institute{
  Department of Physics, Kyoto University, Kyoto 606-8502, Japan
 }

\authorrunning{Takasan, Nakagawa and Kawakami}

\begin{document}

\maketitle


\begin{abstract}
In this paper, we theoretically investigate how laser fields change the nature of topological Kondo insulators(TKIs). By employing a prototypical model of TKIs, we treat the effect of the laser fields with Floquet theory, which gives effective description under high frequency laser fields. We derive the effective model of TKIs under the laser irradiation and discuss its topological properties. We demonstrate a possible realization of Floquet Chern insulators specified by various values of Chern number and reveal how the topological phase changes with increasing the laser light intensity. Furthermore, it is shown that Floquet Weyl semimetals, which have some pairs of Weyl nodes protected topologically, can emerge in the three-dimensional case. We explain how the Weyl nodes are created with varying the strength of the laser field. 

\end{abstract}


%
%


\section{Introduction}

In recent research for new topological phases of matter, topological Kondo insulators (TKIs) have been proposed and have attracted lots of research interest\cite{Dzero2015}. Unlike the other well studied topological insulators, e.g. Bi$_2$Se$_3$, TKIs are strongly correlated electron systems that have robust conducting surface states protected by time-reversal symmetry. Theoretical study by Dzero \textit{et al.}\cite{Dzero2010} opened a novel direction of research for interacting topological matter. From the experimental viewpoint, SmB$_6$ is believed as a strong candidate for TKIs. There are many pieces of evidence supporting that SmB$_6$ has topologically robust surface states at low temperatures, while negative results are also obtained experimentally. Although it is still an important open question to figure out whether the Kondo insulator SmB$_6$ is topological or not, TKIs certainly provide a versatile platform to study topological phases of matter with strong correlation.

Besides intensive studies on topological properties in equilibrium, in recent years, there has been tremendous progress in emergent quantum phases of matter induced  by laser light applications. Especially, it has been intensively investigated what topological phases are induced by laser light. For example, it is known that the application of linearly polarized laser can induce a topological state in a semiconductor quantum well, which is initially in the trivial phase\cite{Lindner2011}. It exemplifies a realization of nonequilibrium states corresponding to $Z_2$ topological insulators, called Floquet topological insulators. There are also other examples showing that the system can be driven to nonequilibrium topological phases in graphene\cite{Kitagawa2011} or cold atomic systems\cite{Nakagawa2013}.

In this study, we theoretically investigate the topological nature of TKIs under laser irradiation. 
TKIs have the unique hybridization term which depends on the (pesudo)spin and the momentum of electrons. It originates from the spin-orbit interaction(SOI). Then a new kind of phenomenon can be induced by the cooperation of the effect of laser light and SOI in the laser-irradiated TKIs. The paper is organized as follows. After introducing the model of TKIs and the method to treat the effect of high frequency laser irradiation, we derive the effective model and discuss its topological properties. We address a realization of the laser-induced topological states of matter such as Chern insulators and Weyl semimetals, which are referred to as Floquet Chern insulators and Floquet Weyl semimetals\cite{Wang2014}.

\section{Model and Method}

\subsection{Model Hamiltonian of topological Kondo insulators}
In order to discuss the laser-induced effects on TKIs , we introduce a simple model\cite{Tran2012}, which consists of the conduction {\it d}-electron band and the localized {\it f}-electron band. This model is a variant of the periodic Anderson model of which hybridization terms have nontrivial dependences on the (pseudo)spin and the momentum due to SOI in \textit{f}-orbit. Assuming the cubic lattice structure similar to $\mathrm{SmB_6}$, which is a strong candidate material for TKIs, the model Hamiltonian reads
\begin{align}
\calH_{\mathrm{TKI}}
&=\sum_{\bm k \in \mathrm{B.Z.}}
\begin{pmatrix}
\bm d_{\bm k}^\dagger \\
\bm f_{\bm k}^\dagger \\
\end{pmatrix}^T
\begin{pmatrix}
\epsilon_d ({\bm k}) \bm 1 &  {\bm V^{(0)}({\bm k}) \cdot \bm \sigma} \\
 {\bm V^{(0)*}({\bm k}) \cdot \bm \sigma} &  \epsilon^{(0)}_f({\bm k}) \bm 1  & \\
\end{pmatrix}
\begin{pmatrix}
\bm d_{\bm k}\\
\bm f_{\bm k}\\
\end{pmatrix}
+U \sum_i n_{i \uparrow}^{(f)} n_{i \downarrow}^{(f)}
\end{align}
with
\begin{align}
\epsilon_d ({\bm k})&=-2 t_d(\cos k_x+\cos k_y +\cos k_z), \\
\epsilon^{(0)}_f ({\bm k})&=\epsilon^{(0)}_f-2 t_f(\cos k_x+\cos k_y +\cos k_z) , \\
\bm V^{(0)}(\bm k)&=i V^{(0)}( a_1 \sin k_x, a_2 \sin k_y, a_3 \sin k_z),
\end{align}
where $\bm 1$ and $\bm \sigma$ are 2$\times$2 identity matrix and Pauli matrices.
We use $(\bm d_{\bm k},\bm f_{\bm k})$ as a short hand notation for 
$(d_{{\bm k} \uparrow},d_{{\bm k} \downarrow},f_{{\bm k} \uparrow},f_{{\bm k} \downarrow})$,
which are annihilation operators in the conduction and the localized band respectively.
Here $\uparrow$ and $\downarrow$ stand for the (pseudo)spin. 
The fact that $\bm \epsilon_d ({\bm k})$ and $\bm \epsilon_f ({\bm k})$ are parity even and $\bm V(\bm k)$ is parity odd guarantees time reversal symmetry of this system and we can define a $Z_2$ topological index.

We incorporate the correlation effect in {\it f}-orbit with a Gutzwiller-type mean field treatment, following Dzero \textit{et al.} \cite{Dzero2010}. We thus assume a renormalized band structure, which results in the mean-field Hamiltonian,\begin{align}
\calH_{\mathrm{MF}}
&=\sum_{\bm k \in \mathrm{B.Z.}} \Psi^{\dagger}(\bm k) \calH_{\mathrm{MF}}(\bm k)\Psi(\bm k)\\
&=\sum_{\bm k \in \mathrm{B.Z.}}
\begin{pmatrix}
\bm d_{\bm k}^\dagger \\
\bm f_{\bm k}^\dagger \\
\end{pmatrix}^T
\begin{pmatrix}
\bm \epsilon_d ({\bm k}) \bm1 &  \bm V({\bm k}) \cdot \bm \sigma\\
\bm V^*({\bm k}) \cdot \bm \sigma &
\bm \epsilon_f({\bm k}) \bm1 & \\
\end{pmatrix}
\begin{pmatrix}
\bm d_{\bm k} \\
\bm f_{\bm k} \\
\end{pmatrix},
\end{align}
where $\Psi(\bm k)=(d_{{\bm k} \uparrow},d_{{\bm k} \downarrow},f_{{\bm k} \uparrow},f_{{\bm k} \downarrow})$, $\bm V= b \bm V^{(0)}=i V( a_1 \sin k_x, a_2 \sin k_y, a_3 \sin k_z)$ and $\epsilon_f({\bm k})= \epsilon_f^{(0)}({\bm k}) +\lambda$. Note that $b$ is a Gutzwiller-type renormalization factor and $\lambda$ represents an energy shift of {\it f}-orbital level by the interaction effect. 

We set the parameters of the model as $t_f = -0.25 t_d$ in this study and vary the parameters $V$, $\epsilon_f$ and $\bm a = (a_1, a_2, a_3)$ according to the situations. Especially we consider two cases of the hybridization parameter $\bm a$ in this study: $\bm a_{\mathrm{2D}} = (2,2,0), \bm a_{\mathrm{3D}} = (2,2,-4)$. Here $\bm a_{\mathrm{2D}}$(with fixing $k_z=0$) and $\bm a_{\mathrm{3D}}$ correspond to a two-dimensional (2D) TKI and a three-dimensional (3D) TKI respectively. 

\subsection{Effective Hamiltonian in Floquet theory}\label{Floquet}

It is known that Floquet theorem is useful for analyzing the system which has time periodic Hamiltonian\cite{Oka2009}. Floquet theorem is, so to speak, ``Bloch's theorem for time direction'': If the Hamiltonian is time periodic $\calH(t)=\calH(t+T)$, the eigenfunction can be written by a product of an exponential function $e^{-i\epsilon t}$ and a time periodic function $u(t)$. $\epsilon$ is called ``pesudo energy'' that is defined in the range of $-\pi/T < \epsilon < \pi/T=\omega/2$. Then, we can define the effective Hamiltonian in Floquet theory, of which eigenvalues are pesudo energy, as
\begin{align}
\calH_{\eff}=\frac{i}{T}\log U(T),
\end{align} 
where the time-evolution operator $U(t)$ is written as $\mathcal{T}\exp \left(-i \int_0^{t} \calH(s)ds \right) $. By definition, the effective Hamiltonian has only the information of $t=0,T,2T,\cdots, nT,\cdots$ and gives ``stroboscopic description" of this system. In this description, the information of the time range of $nT<t<(n+1)T$ is neglected. However, if the time period is short enough, this Hamiltonian tells us the effective behavior of this system; namely it describes the asymptotic behavior in the high frequency limit. Then we can understand the nonequilibrium states in the high frequency limit with the effective equilibrium model. 

There are some methods to calculate the effective Hamiltonian. We adopt a perturbative approach following Kitagawa \textit{et al.}\cite{Kitagawa2011}. With this approach, we can write down the effective Hamiltonian as
\begin{align}
\calH_{\mathrm{eff}} = \calH_0 + \frac{[\calH_{-1},\calH_1]}{\omega}+\calO\left(\left(\frac{A}{\omega}\right)^3\right), \label{Heff}
\end{align}
where $\calH_n =\frac{1}{T} \int^{T/2}_{-T/2}dt \calH(t) e^{i n \omega t}$. The second term in the case of the laser-irradiated systems can be understood as the second order perturbation process of the virtual photon absorption and emission in off-resonant light. We derive and investigate the effective model under the high frequency laser fields with this approach in the following section.

\section{Results and Discussion}

\subsection{Derivation of the effective model}

We derive the effective model of TKIs under the laser field $\bm A = (A_x \sin \omega t,A_y \sin( \omega t + \phi),0)$, where $\phi$ is the polarization angle of the laser light. We focus on $\phi=\pi/2$(circularly polarized) case in this study. In this case, the circularly polarized laser light breaks the time-reversal symmetry of the original system. The symmetry class is changed by the laser irradiation and thus it is expected that the application of the laser light makes the nature of the system dramatically different. The effect of the laser light is taken into account by minimal coupling $\bm k \rightarrow \bm k -\bm A$, which corresponds to introducing a Peierls phase $t_{ij} \rightarrow t_{ij} e^{i \phi_{ij}} $ in the real space representation. Using $\calH(t)=\sum_{\bm k \in \mathrm{B.Z.}} \Psi^{\dagger}(\bm k) \calH_{\mathrm{MF}}(\bm k - \bm A (t))\Psi(\bm k)$, we calculate the effective Hamiltonian with the method in Sec.\ref{Floquet}. The explicit form of the effective Hamiltonian obtained for $\phi=\pi/2$ is 
\begin{align}\label{Hcir}
\calH_{\eff}
&=\sum_{\bm k \in \mathrm{B.Z.}}
\begin{pmatrix}
\bm d_{\bm k}^\dagger \\
\bm f_{\bm k}^\dagger \\
\end{pmatrix}^T
\begin{pmatrix}
\tilde{\epsilon}_d ({\bm k})\bm1 + \Phi_B({\bm k})  \sigma_z&  \left( \widetilde{\bm V}({\bm k}) + \Phi(\bm k) \right) \cdot \bm \sigma \\
\left( \widetilde{\bm V}^*({\bm k}) + \Phi^*(\bm k) \right)\cdot \bm \sigma  &
\tilde{\epsilon}_f({\bm k})\bm1 + \Phi_B({\bm k})  \sigma_z& \\
\end{pmatrix}
\begin{pmatrix}
\bm d_{\bm k} \\
\bm f_{\bm k} \\
\end{pmatrix},
\end{align}
where
\begin{align}
\tilde{\epsilon}_d ({\bm k})&=-2 t_d(J_0(A_x) \cos k_x+J_0(A_y)\cos k_y +\cos k_z), \\
\tilde{\epsilon}_f ({\bm k})&=\epsilon_f - 2 t_f(J_0(A_x) \cos k_x+J_0(A_y) \cos k_y +\cos k_z ),\\
\widetilde{\bm V}(\bm k)&=i V( a_1 J_0(A_x) \sin k_x, a_2 J_0(A_x) \sin k_y, a_3 \sin k_z),
\end{align}
\begin{align}
\Phi_B(\bm k) &= \frac{2V^2 J_1(A_x)J_1(A_y) }{\omega}(a_1a_2^*+a_2a_1^*) \cos k_x \cos k_y,\\
\bm \Phi(\bm k)&=iV \frac{4(t_d- t_f)J_1(A_x)J_1(A_y) }{\omega}(a_1 \cos k_x \sin k_y, -a_2 \sin k_x \cos k_y,0),
\end{align}
and $J_n(x)$ is the $n$-th Bessel function.

We can see some effects of the laser light in eq.(\ref{Hcir}). 
i) \textit{Photo-renormalization effect}: 
We can see the renormalization of the hopping integrals $t_d, t_f$ and the hybridization term $V$ by the 0-th Bessel function. Regarding hopping integrals, this effect is nothing but ``dynamical localization"\cite{Dunlap1986}, which claims that the hopping integral of electrons is reduced by the laser light because the frequency is too high for the motion of electrons to follow the AC electric fields of the laser light. The effect on the hybridization term can also be understood as an analog of dynamical localization since TKIs have only inter-site hybridization unlike the Kondo insulators described by the conventional periodic Anderson model.
ii) \textit{Photo-induced hybridization}: 
From the commutator term in eq.(\ref{Heff}), we obtain new hybridization terms $\bm \Phi (\bm k)$. They are derived from spin-dependent hybridization terms originating from SOI. The forms of these terms reflect the $\pi/4$-rotational symmetry ($k_x \rightarrow k_y$, $k_y \rightarrow -k_x$) since the laser light has the same symmetry. 
iii) \textit{Photo-induced ``magnetic" field}: 
$\Phi_B (\bm k) $ are derived from the commutator of the $x$ and $y$ components of hybridization terms and play a role of the Zeeman coupling to (pesudo)spins. Then we can regard them as momentum-dependent ``magnetic" fields originating from SOI. These terms break time-reversal symmetry in a similar way to usual magnetic fields. In the following sections, we will see that the effect iii) particularly plays a vital role in inducing topological phase transitions.

\subsection{Topological invariant}

We investigate the topological nature of the effective model derived in the previous section. In general, the topological nature is characterized by a topological invariant, which is robust against any local perturbations without breaking the relevant symmetry of the system, e.g. time reversal symmetry or inversion symmetry. For this purpose, we use the Chern number $C$, which is a topological invariant defined in two-dimensional systems. The Chern number can be calculated as \cite{Volovik_book}  
\begin{align}
C = \frac{1}{24 \pi^2}\int^{\infty}_{-\infty} d\varepsilon \int_{\mathrm{B.Z.}} dk_x dk_y \mathrm{Tr}[\epsilon^{\mu \nu \rho} (G \partial_\mu G^{-1})(G \partial_\nu G^{-1})(G \partial_\rho G^{-1})], 
\end{align}
where the Green function $G^{-1}(\varepsilon, \bm k)=i \varepsilon - \calH_{\eff}(\bm k)$ and $\mu,\nu,\rho$ run through $\varepsilon, k_x, k_y$.
As for 2D systems, it is known that the Chern number can characterize integer quantum Hall states or Chern insulating states. Moreover, topological characterization by the Chern number can be applied to 3D systems. Weyl semimetals, which have an even number of topologically protected gapless nodes (Weyl nodes), can be also characterized by the Chern number, which is calculated on a 2D plane in the 3D Brillouin zone. If we choose two planes which hold a Weyl node between them, we obtain the different Chern numbers from each plane. This difference corresponds to the topological number which the Weyl node has.

\subsection{2D case : Floquet Chern insulator}

\begin{figure}[htbp]
\begin{center}
\includegraphics[width=8cm]{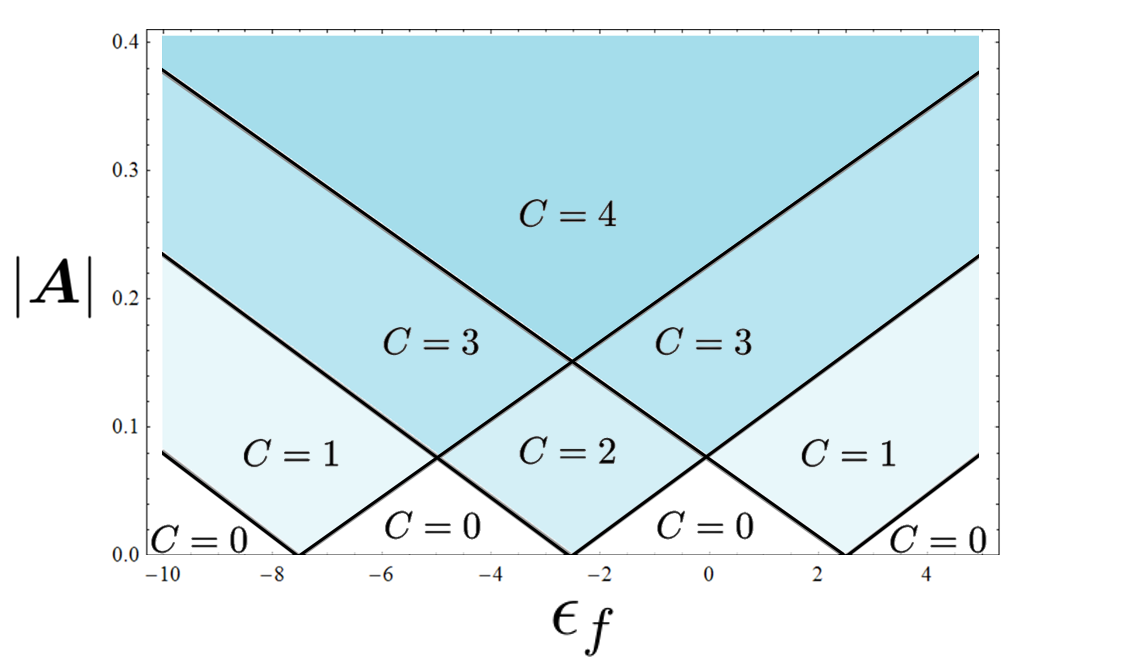}
\caption{Topological phase diagram of our effective model in the 2D case. The parameter is taken as $V= 2.0 t_d$. We can realize various phases that have different Chern numbers with varying the laser intensity $|\bm A|$ or the energy level of {\it f}-orbit $\epsilon_f$. }
\label{phase diagram}
\end{center}
\end{figure}

First, we study the 2D case in which we use the parameters ${\bm a}_\mathrm{2D}= (2,2,0)$ and set $k_z=0$. We can calculate the Chern number of this system since the system is 2D and gapped. For several values of $|\bm A|$ and $\epsilon_f$, we numerically evaluate the non-trivial Chern numbers. From this calculation, we obtain the topological phase diagram of this system, which is shown in Fig.\ref{phase diagram}. This phase diagram shows that the system can be Floquet Chern insulator (Quantum Hall insulator) phases, which have non-trivial Chern numbers. It is seen that they have some chiral edge modes corresponding to the Chern numbers of each phase. There are intriguing phases that have high Chern numbers $C>1$ and $|C|$ chiral edge modes. 

\begin{figure}[htbp]
\begin{center}
\includegraphics[width=14cm]{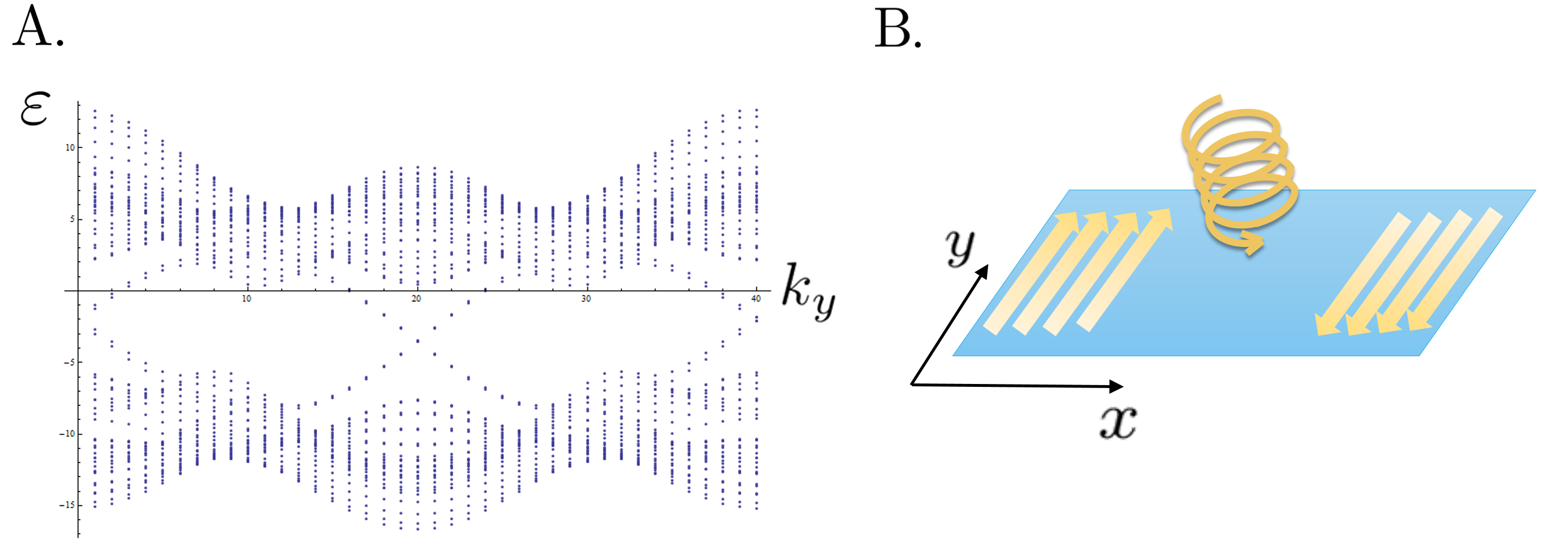}
\caption{(A) Band structure of Floquet Chern insulator calculated in OBC. The parameters are taken as $|A|=0.3$ ,$\epsilon_f=-4.0 t_d$, and $V= 3.0 t_d$. (B) Schematic picture which describes the propagation of the four chrial edge modes when the circularly polarized laser irradiates the 2D system. We assume that the system size is finite in $x$-direction and infinite in $y$-direction. }
\label{Chern_band}
\end{center}
\end{figure}

Furthermore, we directly confirm the existence of the chiral edge modes by numerical calculation of the band structure. We calculate the band structure in the open boundary condition (OBC) in $x$-direction and the periodic boundary condition (PBC) in $y$-direction. By this calculation, we obtain the band structure of Floquet Chern insulator that has $C=4$ in Fig.\ref{Chern_band}A. We can see from the figure that there are eight modes on the edges in $x$-direction (Four nearly degenerate edge modes are at $k=0$ and the other four edge modes are at $k=\pm \pi $). Further investigation of the eigenfunctions of the edge modes elucidates that the four edge modes which propagate in the same direction are localized in the same side of the edges. We thus confirm that there are four ``chiral'' edge modes on each side, as shown in Fig.\ref{Chern_band}B. We can control the number of these edge modes by tuning the laser light intensity $|\bm A|$.

Here, we remark difference between our study and the previous researches.
It is known that photo-induced quantum Hall effect happens in graphene under the circularly polarized laser light\cite{Oka2009}.
This result is similar to our work in that the circularly polarized laser enables to acquire the non-trivial Chern number and induces the chiral edge modes. However, in the previous work, the effect of breaking the time reversal symmetry is induced by photo-assisted hopping between the next-nearest neighbors, whereas the photo-induced ``magnetic field'' originating from SOI breaks the time reversal symmetry in our model. There is not SOI in graphene and thus our proposal is different from theirs in the way to break the time reversal symmetry. 

\subsection{3D case : Floquet Weyl semimetal}

\begin{figure}[htbp]
\begin{center}
\includegraphics[width=14cm]{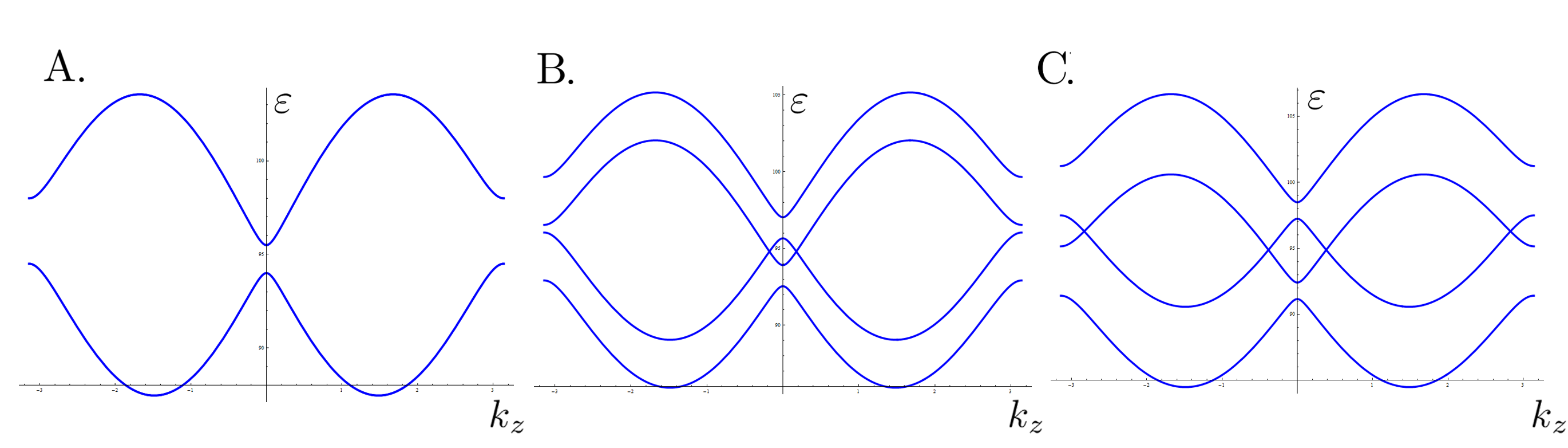}
\caption{Band structures for $k_z$ in the 3D case ($k_x$ and $k_y$ are fixed to zero). They are calculated in PBC for some values of laser intensity when $\epsilon_f$ and $V$ are respectively fixed to $-6.0t_d$ and $2.0 t_d$. (A),(B), and (C) correspond to $|\bm A| = 0, 0.1 t_d, 0.2 t_d$ respectively. (A) represents the band structure of a TKI. In (B) and (C), we can see the Weyl nodes, which suggest that a Floqeut Weyl semimetal is realized.}
\label{Weyl_band}
\end{center}
\end{figure}

Next, we study the 3D case in which we use the parameters ${\bm a}_\mathrm{3D}= (2,2,-4)$. We numerically calculate the bulk band structure using the effective Hamiltonian, and show how the band structure changes with increasing the laser intensity in Fig.\ref{Weyl_band}($k_x$ and $k_y$ are fixed to zero).  We show the initial state before the laser irradiation in Fig.\ref{Weyl_band}A, and the band structure when the laser intensity is finite in Fig.\ref{Weyl_band}B and \ref{Weyl_band}C. Weyl nodes are first created at $\bm k=(0,0,0)$ and next at ${\bm k} = (0,0,\pi)$, and some pairs of Weyl nodes can be seen in Fig.\ref{Weyl_band}B and \ref{Weyl_band}C. These results show the possibility that a Floquet Weyl semimetal\cite{Wang2014} is realized in this system. However, there is another possibility that these nodes are made by accidental band touching. To figure out this point, we examine the Chern number for several values of $k_z$ and find that the Chern number indeed changes if the $k_x$-$k_y$ plane (fixing $k_z$) goes over the Weyl nodes when $k_z$ varies. This confirms that these nodes are topologically protected, thus ensuring the system to be a Weyl semimetal. 

Weyl semimetals receive a lot of attention recently as an exotic gapless topological phase of matter. There are interesting phenomena theoretically predicted in Weyl semimetals, such as the anomalous Hall effect\cite{Yang2011} and the chiral magnetic effect\cite{Fukushima2008}. Unfortunately, there are only few experimental observations\cite{Huang2015} of Weyl semimetals, and therefore it is desired to experimentally realize Weyl semimetals. 
An interesting feature of our proposal to realize Weyl semimetals in a nonequilibrium state is that we can control the number or position of Weyl nodes with varying the laser light intensity (see in Fig.\ref{Weyl_band}C).
 This may allow us to control and observe characteristic surface states inherent in Weyl semimetals, called \textit{Fermi arc}\cite{Wan2011} which connects the different Weyl nodes. 

\section{Summary}

We have examined how laser fields change the nature of TKIs. By employing a prototypical model of TKIs, we have treated the effect of the laser field with the effective Hamiltonian in Floquet theory. We have derived the effective model of TKIs under the laser irradiation and  have revealed the effects of laser light. There are the effects originating from the spin- and momentum-dependent hybridization terms that are unique to TKIs. Then we have shown that the effects of laser light induce Floquet Chern insulators with various values of the Chern number and have elucidated how the topological phase transitions occur. Moreover, we have addressed a possible realization of Floquet Weyl semimetals, which have some pairs of topologically protected Weyl nodes.

In this paper, we have not discussed the important effects due to the laser-irradiation, i.e. the effect of heating, which should be crucial to realize our proposal in real experiments. This issue is now under consideration.

This work is supported by KAKENHI (Grants No. 25400366 , No. 14J01328, and No. 15H05855). M.N. thanks JSPS for the support from a Research Fellowship for Young Scientists.

\bibliography{ref.bib}

\end{document}